\documentclass[]{interact}

\usepackage{xcolor}
\usepackage{epstopdf}
\usepackage[caption=false]{subfig}

\usepackage[numbers,sort&compress,merge]{natbib}
\bibpunct[, ]{[}{]}{,}{n}{,}{,}

\theoremstyle{plain}

\theoremstyle{definition}

\theoremstyle{remark}

\begin{document}

\articletype{ARTICLE}

\title{Diffusion limited aggregation, resetting and large deviations of Brownian motion}

\author{
\name{Uriel Villanueva-Alcal\'a\textsuperscript{a}, Jos\'e R. Nicol\'as-Carlock\textsuperscript{a} and Denis Boyer\textsuperscript{a}\thanks{Email: boyer@fisica.unam.mx}}
\affil{\textsuperscript{a} 
Instituto de Física, Universidad Nacional Autónoma de México, Mexico City 04510, Mexico}
}

\maketitle

\begin{abstract}
Models of fractal growth commonly consider particles diffusing in a medium and that stick irreversibly to the forming aggregate when making contact for the first time. As shown by the well-known diffusion limited aggregation (DLA) model and its generalisations, the fractal dimension is sensitive to the nature of the stochastic motion of the particles. Here, we study the structures formed by finite-lived Brownian particles, {\it i.e.}, particles constrained to find the aggregate within a prescribed time, and which are removed otherwise. This motion can be modelled by diffusion with stochastic resetting, a class of processes which has been widely studied in recent years. In the short lifetime limit, a very small fraction of the particles manage to reach the aggregate. Hence, growth is controlled by atypical Brownian trajectories, that move nearly in straight line according to a large deviation principle. In $d$ dimensions, the resulting fractal dimension of the aggregate decreases from the DLA value and tends to 1, instead of increasing to $d$ as expected from ballistic aggregation. In the zero lifetime limit one recovers the non-trivial model of \lq\lq aggregation by the tips" proposed long ago by R. Jullien [J. Phys. A: Math. Gen. {\bf 19}, 2129 (1986)].
\end{abstract}


\section{Introduction}

Resetting processes have attracted a considerable attention during the last decade in the field of non-equilibrium statistical physics \cite{evans2020stochastic}. Let us consider an arbitrary stochastic process evolving in time and which is interrupted to be reset to its initial state, from which it restarts anew. If the operation is repeated again and again at random times, a resetting process is obtained. A paradigmatic model is given by a Brownian particle which is instantaneously reset at a constant rate to its initial position, while it diffuses freely between two consecutive resetting events \cite{evans2011diffusion}. Resetting violates detailed balance and gives rise to a wealth of new phenomena, such as non-equilibrium steady states \cite{evans2011diffusion,evans2011bdiffusion} or dynamical transition in their temporal relaxation \cite{majumdar2015dynamical}. Resetting can also expedite the time needed by an arbitrary process to reach a certain position (or state) for the first time \cite{evans2011diffusion,evans2011bdiffusion,reuveni2016optimal,chechkin2018random,pal2017first,eliazar2020mean,maso2019transport}, a property which is particularly interesting for understanding the efficiency of enzymatic reactions \cite{reuveni2014role,rotbart2015michaelis,pal2019landau} and for applications to random search problems in ecology \cite{kusmierz2014first, campos2015phase} or network science \cite{riascos2020random}.

Stochastic resetting in interacting particle systems has also been the subject of several studies (see \cite{nagar2023stochastic} for a review).
Resetting a spatially extended system to a particular configuration has important consequences on its steady state and dynamical properties. Some classical models have been revisited under resetting protocols in various contexts: fluctuating interfaces \cite{gupta2014fluctuating}, populations genetics \cite{kang2022evolutionary}, the dynamics of prey-predator systems \cite{mercado2018lotka,da2021diffusion,evans2022exactly}, 
non-conserving zero-range processes \cite{grange2020non}, the 
Ising model \cite{magoni2020ising}, directed polymers in random force fields \cite{grange2020susceptibility}, exclusion processes \cite{miron2021diffusion} or
binary aggregation with constant kernel \cite{grange2021aggregation}, for instance. Notably, the simultaneous resetting of many independent particles creates non-equilibrium steady states with strong correlations \cite{biroli2023extreme}. 

In this article, we consider the impact of resetting processes on fractal growth.
Fractal growth phenomena are out of equilibrium processes that produce disordered spatial systems with self-similar features that are ubiquitous in nature \cite{meakin1998fractals}. Among these, Laplacian growth stands as a paradigmatic model that reproduces and describes the fractal patterns observed in electrodeposition, viscous fingering, colonies of bacteria, dielectric breakdown, vascular systems and cities (see \cite{sander2000diffusion, Sander2011fractal} for reviews). In the Laplacian framework, as introduced in the on-lattice dielectric breakdown model (DBM) \cite{niemeyer1984fractal}, the growth probability of a point on the cluster interface is given by $p \propto |\nabla\phi|^\eta$. Here, $\phi$ is associated to the potential energy of the growing surface, and $\eta$ is a parameter that leads to the formation of compact clusters ($\eta=0$), dendritic fractals ($\eta\sim 1$), and linear structures ($\eta>4$). Notably, the solution to $\eta=1$ leads to a fractal that is structurally equivalent to the classical on-lattice diffusion-limited aggregation (DLA) model \cite{witten1981diffusion,witten1983diffusion, nicolas2019universal}, a very simple model that aims to replicate the growth of clusters limited by the diffusion of particles performing random walks or Brownian motion before aggregation. 

The DLA model has been subject to multiple extensions in order to study the effect of the particles dynamics on the cluster morphology upon aggregation, such as varying the random-walk length \cite{huang1987effects}, having the random walkers perform L\'evy flights \cite{meakin1984levy}, imposing drifts and angular biases \cite{meakin1983effects, huang2001particle}, or by setting attractive/repulsive particle-cluster interactions \cite{block1991aggregation, nakagawa1992extended}. However, in all those extended models (see also the review by Meakin \cite{meakin1998fractals}), the random walkers are considered to have an infinite lifetime, that is, a particle moves until reaching aggregation unless it wanders beyond an \textsl{ad-hoc} killing radius far from the main cluster, in which case, a resetting criteria is implemented in order to speed-up the aggregation process. In fact, this mechanism has no consequence on the overall expected growth. The aim of the present article is to explore the effects that other resetting protocols imposed on the motion of the particles can have on the morphology of the clusters.
Physically, resetting provides a way of modelling particles that have a finite lifetime, {\it i.e.}, that are removed from the system if they do not reach the aggregate before a certain time, and remain permanently aggregated otherwise. A finite lifetime can be caused by diverse mechanisms, such as particle denaturation, resulting in the loss of its binding ability, or unbinding from the substrate, or trapping by impurities, for instance.

This paper is organised as follows. Section 2 describes DLA processes with particles subject to resetting. More specifically, Section 2.1 exposes the two resetting protocols that are used here, whereas Section 2.2 presents a study of the large resetting rate limit (short lifetimes) based on a theory of large deviations for constrained diffusive systems. Results obtained from numerical experiments are shown in Section 3: a comparison with ballistic aggregation and measurements of the fractal dimension, in Section 3.1 and 3.2, respectively. We conclude in Section 4.

\section{DLA models with finite-lived diffusing particles}

In this section, we introduce two models that extend DLA to the case where the diffusing particles have a finite lifetime, whereas the particles that belong to the aggregate are permanent and fixed in space. To gain a qualitative insight on the effect of mortality on the fractal structures that are formed, we subsequently analyse these models in the limit of vanishing lifetime, a case which is tractable by making use of path integral representations of Brownian motion.

\subsection{Two resetting protocols}

Let us consider a two-dimensional square lattice with unit spacing. In ordinary DLA, a seed particle is fixed at the origin site and a diffusive particle starts from an initial position which is randomly and uniformly distributed on a circle of given radius $R\gg1$, centred at the origin. The particle performs an unbiased random walk on the lattice until it reaches a nearest-neighbour (n.n.) site of the seed particle for the first time, where it stops to form part of the aggregate. The process is repeated with another particle, starting from a new random position on the circle, and so on iteratively. Each particle irreversibly binds to the aggregate as soon as it occupies an empty site neighbouring one of the previously aggregated particles. In practice, if the distance between the diffusing particle and the origin becomes larger than, say, $2R$, the particle is discarded and another one is launched from a new position on the circle \cite{nicolas2016fractality}.

We propose a modification of this model where the random walk acquires a finite lifetime. Let us consider the process in discrete time: at each time-step $t\rightarrow t+1$, with probability $1-p_d$ the particle  performs a random walk step to one of its n.n. sites or dies with the complementary probability $p_d$, {\it i.e.}, is removed from the system.
A surviving particle binds to the aggregate as in original DLA and, importantly, the particles belonging to the aggregate become permanent (do not die) and are fixed in space (there is no cluster reconfiguration). As before, once a particle is aggregated, another one is launched from a new random position on the circle of radius $R$.

When a particle dies, it can no longer reach the aggregate and one needs to specify what happens with the next particle. 
Two variants are considered here:

\begin{itemize}
  \item Model A: If the diffusing particle dies, a new random initial position is chosen on the launching circle for the following particle.
  \item Model B: If the diffusing particle dies, the following particle starts from the same initial position on the circle than the particle that just died.
\end{itemize}
The above rules actually describe diffusion processes with stochastic resetting. Model A is equivalent to consider a same particle whose motion, at each time step, is interrupted with probability $p_d$ and reset at the next time-step to a new position on the launching circle; whereas in Model B, the particle is reset to its starting position (restart), a case of particular interest in the original resetting model of \cite{evans2011diffusion} and in many subsequent works. In both cases, when the particle is finally aggregated, a new initial position is chosen for the next particle. As shown further, these two variants generate quite different structures.

Due to the Markov nature of the dynamics, the probability distribution of the random walk lifetime $t_{\ell}$, with  $t_{\ell}=0,1,2,...$, is given by $(1-p_d)^{t_{\ell}}p_d$  (in the absence of aggregation). In the limit $p_d\ll 1$ of interest in numerical simulations, $ t_{\ell}$ is thus exponentially distributed with average $\langle t_{\ell}\rangle=1/p_d$, {\it i.e.}, $P(t_{\ell})\simeq p_de^{-p_dt_{\ell}}$. 
It is convenient in the following to rewrite $\langle t_{\ell}\rangle$ or $p_d$ in terms of the typical number of steps $R^2$ needed by the random walk to be at a distance $R$ from its starting point. We define the re-scaled death/resetting probability $\lambda$ through the relation,
\begin{equation}\label{pd}
p_d=\frac{\lambda}{R^2},
\end{equation}
 or $\lambda=R^2/\langle t_{\ell}\rangle$. One can also consider the Brownian limit, where Gaussianly distributed displacements with variance $\sigma^2$ are generated in small time intervals of duration $\Delta t$, with the diffusion coefficient of the particle given by $D=\sigma^2/(4\Delta t$) in $2d$. During $\Delta t$, the probability that the particle is reset is $r\Delta t$, where $r$ is the resetting rate. In this case $\langle t_{\ell}\rangle=1/r$. In analogy with Eq. (\ref{pd}), we define,
\begin{equation}\label{r}
r=\frac{4D\lambda}{R^2},
\end{equation} 
where $\lambda$ now represents the
adimensional resetting rate and $R^2/(4D)$ is the typical diffusion time from the launching circle to the aggregate. In the lattice random walk $D=1/4$, $\Delta t=1$ and Eq. (\ref{pd}) is recovered from Eq. (\ref{r}). Whereas 
$0\le p_d\le 1$, the resetting rate can be any positive real number in the continuous time limit $\Delta t\to0$. Depending on whether time is discrete or continuous, we will use Eq. (\ref{pd}) or Eq. (\ref{r}) to refer to the adimensional parameter $\lambda$, respectively.

\subsection{Geometrical optics of constrained Brownian motion}

In ordinary DLA ($\lambda=0$), the trajectories followed by the particles until aggregation are typical realisations of the random walk. Solving the Laplace equation in discrete space with an absorbing condition at the aggregate boundary allows in principle to obtain the aggregation probability at each point \cite{witten1983diffusion,turkevich1985occupancy}. If $\lambda\gg 1$, however, the situation is quite different, as the dynamics become highly constrained by the lifetime: since the latter is much shorter than the typical diffusion time, only a small fraction of the particles actually reaches the aggregate. The corresponding trajectories are no longer representative of free motion but are pushed into a large deviation regime instead.

Let us adopts a continuous space-time description where the diffusing elements are Brownian particles. To evaluate in the large $\lambda$ limit the aggregation probability at a position $x$ on the aggregate boundary, we can resort to a very simple yet powerful method, the optimal fluctuation method  \cite{basnayake2018extreme,meerson2019large}. In the case of Brownian motion, this method reduces to geometrical optics and, somehow surprisingly, has been shown to be equivalent to exact results in the large deviation regime in a variety of rather complicated geometries \cite{meerson2019geometrical}. We recall this method below along the lines of \cite{meerson2019geometrical}.

One starts by writing the probability of occurrence of a particular Brownian path $X(t)$ with $0\le t\le t_f$ as \cite{risken1989fokker, zinn2004path,majumdar2005brownian}
\begin{equation}
P[\{X(t)\}]\propto e^{-\frac{1}{4D}\int_0^{t_f}\dot{X}(t)^2 dt},
\end{equation}
up to a proportionality constant and where $t_f$ is an arbitrary time.
We denote $W(x,t_f|x_0)=\langle \delta(X(t_f)-x)\rangle$ as the probability density of presence around $x$ at time $t_f$ given the initial condition $x_0$. This quantity takes the form of a path integral,
\begin{equation}\label{path}
W(x,t_f|x_0)\propto \int {\cal D}[X(t)]e^{-\frac{1}{4D}\int_0^{t_f}\dot{X}(t)^2 dt}
\end{equation}
where the sum runs over all paths with $X(0)=x_0$ and $X(t_f)=x$, and where ${\cal D}[X(t)]$ denotes the integration measure. The most probable trajectory is the one that minimise the Wiener action $S=\frac{1}{4D}\int_0^{t_f}\dot{X}(t)^2 dt$. As shown below, when the time $t_f$ becomes very short ($x$ being fixed) the most probable trajectory dominates the sum (\ref{path}) and a saddle-point evaluation can be performed: 
\begin{equation}\label{saddle}
W(x,t_f|x_0)\propto e^{-S^{\ast}},
\end{equation}
where $S^{\ast}$ is the minimum action. To see this, one first performs the functional minimisation of $S$, which leads to $\ddot{X}(T)=0$, or the speed $|\dot{X}(t)|$ must be constant. Denoting the total length of the path as ${\cal L}$, one deduces
\begin{equation}\label{sstar}
S=\frac{1}{4D}\int_0^{t_f}\left(\frac{{\cal L}}{t_f} \right)^2 dt=\frac{{\cal L}^2}{4Dt_f}.
\end{equation}
The minimum action $S^{\ast}$, and therefore $W$ in Eq. (\ref{saddle}), is obtained from finding the minimal path length ${\cal L}^{\ast}$ between $x_0$ and $x$ under the specific geometrical constraints of the problem under study, in analogy with geometrical optics.
One actually checks from Eqs. (\ref{saddle})-(\ref{sstar}) that at very small $t_f$, any path with ${\cal L}>{\cal L}^*$ has a much lower probability and  contributes little to the path integral. 
This method is therefore suitable to study the short time behaviour of Brownian motion problems with constrains, and can be applied to Models A and B in continuous space and time when the lifetime is very small, {\it i.e.}, in the limit $\lambda\rightarrow\infty$. In other words, we now considers that the resetting rate in Eq. (\ref{r}) tends to infinity.

\begin{figure}
\centering
\includegraphics[width=0.49\linewidth]{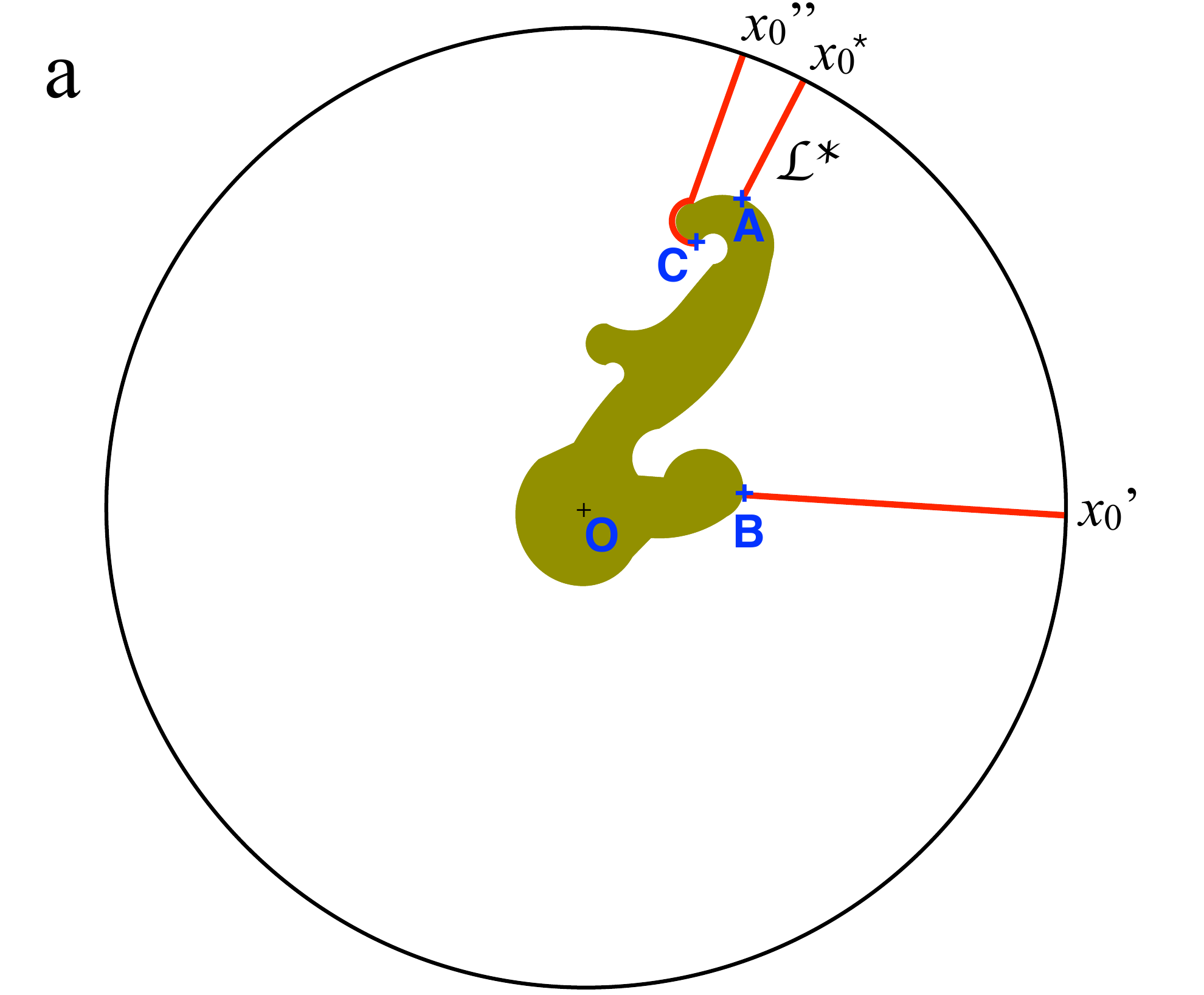}
\includegraphics[width=0.49\linewidth]{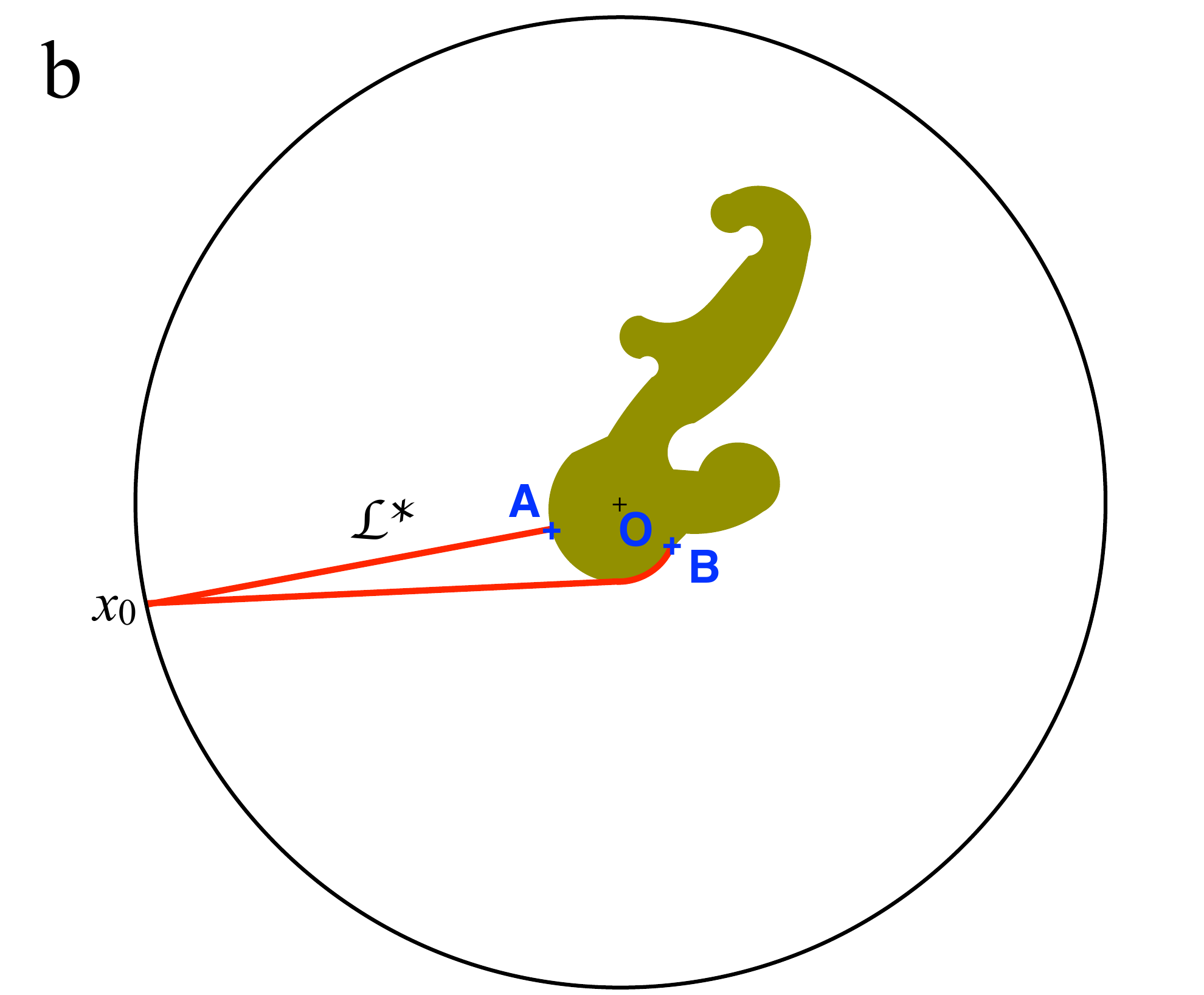}
\caption{Sketch of an aggregate (in green) in continuous space. (a) Model A: the red lines indicate the shortest paths to reach the circle from the boundary points $A$, $B$, and $C$, without crossing the aggregate.  (b) Model B: the shortest paths to two points $A$ and $B$ from a same $x_0$ on the circle. In each case, the shortest of all paths has length ${\cal L}^*$.} \label{geom}
\end{figure}

In the case of Model A, the initial condition is variable and therefore part of the optimisation process: the most likely cluster site for aggregating the next particle is simply the one that is the closest to the launching circle (or the farthest from the seed). It is reached by a straight line trajectory starting at $x_0^*$ and arriving at $A$ in Fig. \ref{geom}a. Any trajectory starting from another point $x_0'$ of the circle and reaching the closest aggregate site ($B$) covers a larger distance, with a aggregation probability $p_B\propto e^{-{\cal L}^2/(4Dt_f)}\ll p_A$.
Notice that the calculation of the shortest path length ${\cal L}$ between an arbitrary boundary point ($C$) and the circle is not straightforward: this path goes from $C$ to some point $x_0^{\prime\prime}$  (Fig. \ref{geom}a) without passing through the aggregate and results  not to be a straight line in general.

In the case of Model B, the initial position $x_0$ is fixed and the optimisation is performed over the boundary sites only. Consequently, the next particle sticks to the boundary site $A$ that is the closest to $x_0$, as indicated in Fig. \ref{geom}b. We also display a shortest but 
sub-optimal path to another point $B$, composed of a straight line and a part closely following the boundary.

\begin{figure}
\centering
\includegraphics[width=0.32\linewidth]{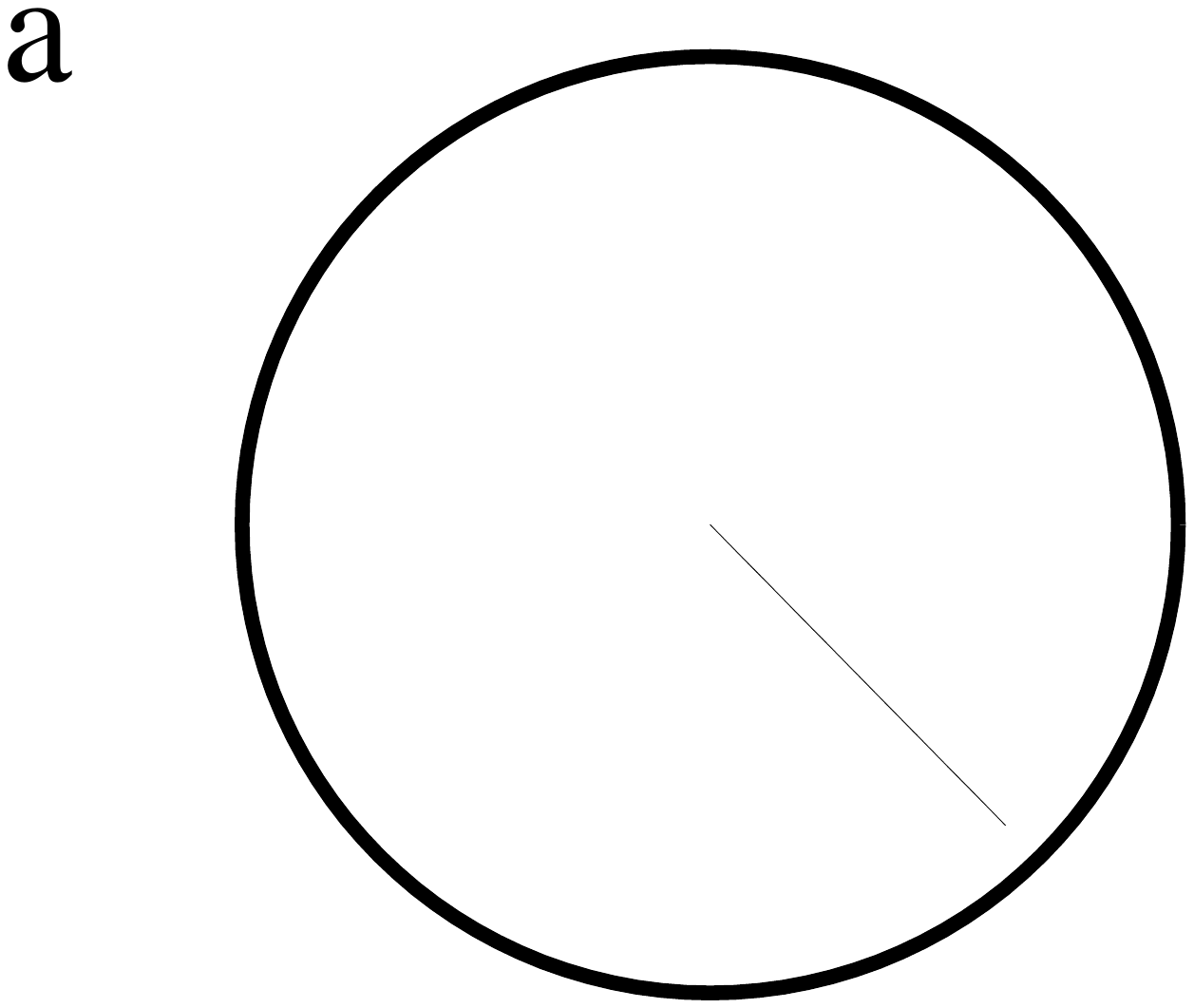}
\includegraphics[width=0.32\linewidth]{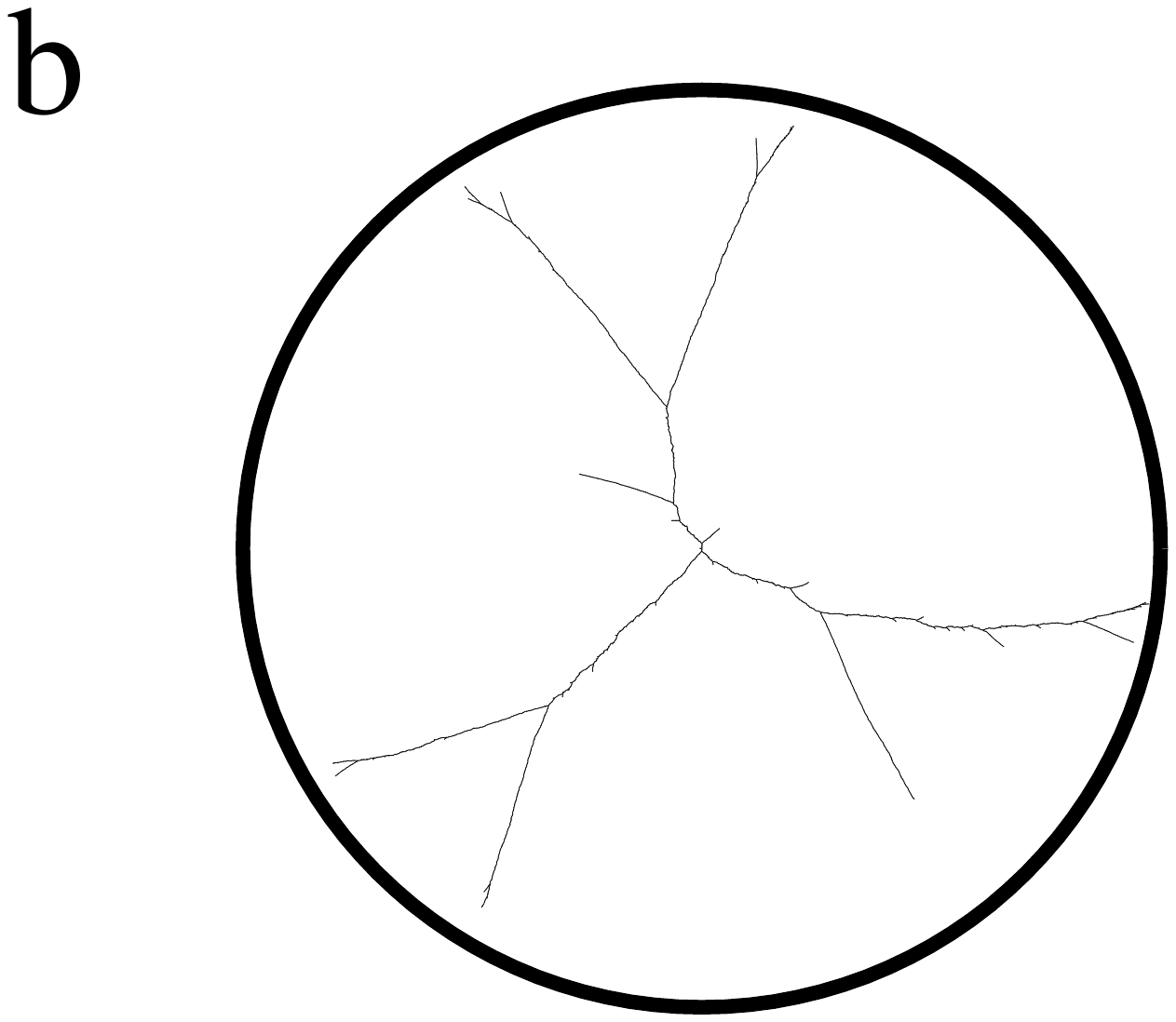}
\includegraphics[width=0.32\linewidth]{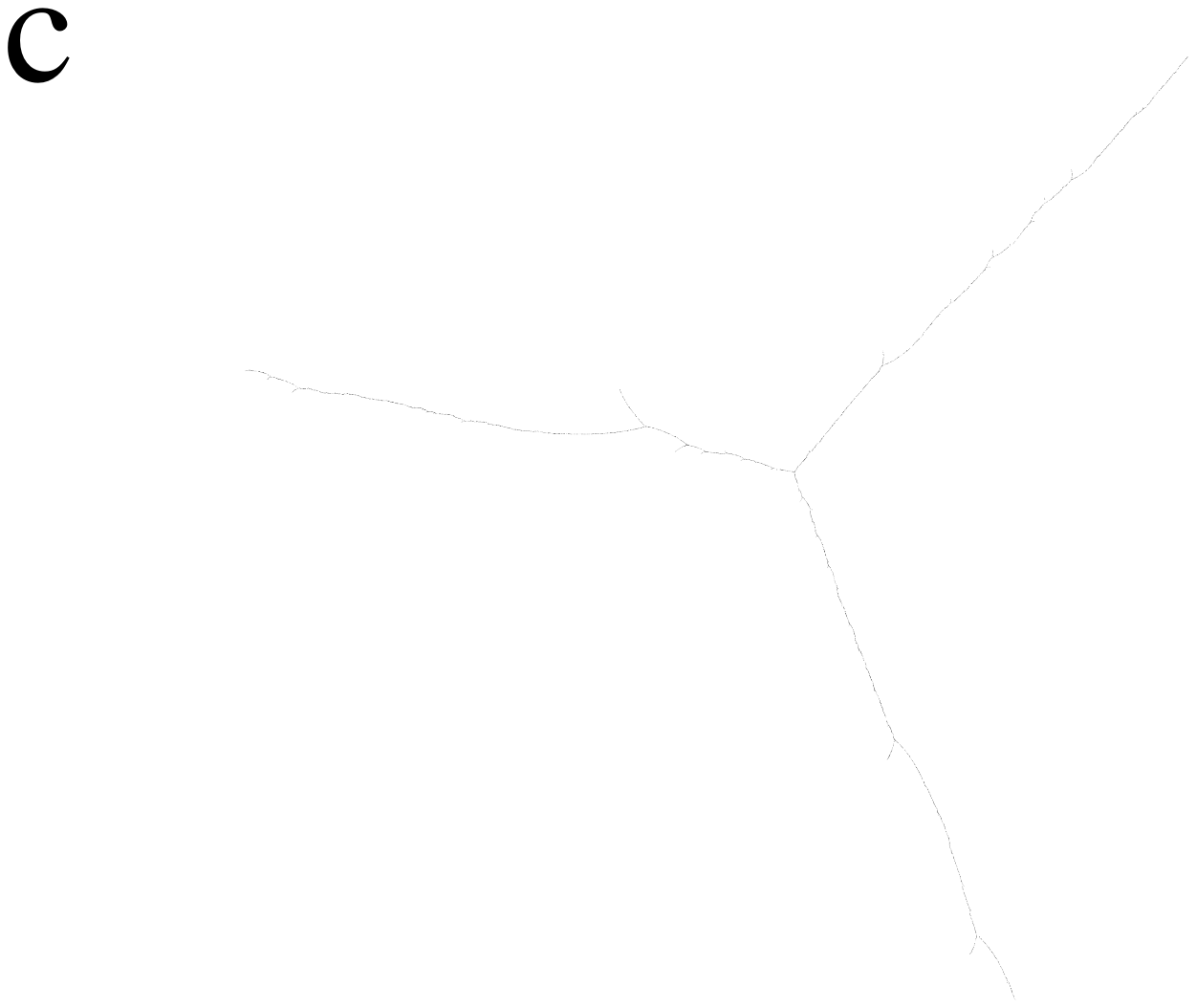}
\caption{Two-dimensional clusters produced in the $\lambda=\infty$ limit of Model A (a) and B (b), with a finite launching radius $R=1000$ and particles of diameter $1$. Space is continuous. (c) Limit $R=\infty$ of case (b). The number of particles in the aggregates is $N=900$, $6000$ and $5\ 10^5$, respectively.} \label{lambdainfty}
\end{figure}

In summary, our two models can be rephrased in the large $\lambda$ limit as follows. Let us consider spherical particles (disks in $2d$) of unit diameter and let us centre the seed at the origin $O$.
\begin{itemize}
  \item Model A ($\lambda=\infty$): Choose a randomly oriented axis $\hat{u}_0$ passing through $O$; add the centre of the second particle on this axis at unit distance of $O$. The centres of the forthcoming particles are added one after another on $\hat{u}_0$, each one at unit distance from the preceding particle.
  \item Model B ($\lambda=\infty$): For adding a new particle, choose a random position $x_0$ on the launching circle, determine the closest aggregate particle $A$, with centre at $x_A$; place the centre of the new particle at distance $1$ from $x_A$ along the direction $(x_A,x_0)$.
\end{itemize}
Trivially, Model A ($\lambda=\infty$) generates a deterministic straight line, see Fig. \ref{lambdainfty}a. The only source of stochasticity is the choice of the orientation when placing the second particle. As the launching circle plays no role, one can also set $R=\infty$. On the other hand, the structures generated by Model B ($\lambda=\infty$) are ramified and far less trivial, as shown by a typical numerical simulation in Fig. \ref{lambdainfty}b. In particular, as the tips of the aggregate approach the circle, more branching occurs.

It is natural to further consider the \lq\lq thermodynamic limit" $R\rightarrow\infty$ of Model B ($\lambda=\infty$), which allows us to get rid of edge effects. The previous rules can be slightly modified to meet this limit:
\begin{itemize}
  \item Model B ($\lambda=\infty$, $R=\infty$): 
  For adding a new particle, choose an axis with random orientation $\hat{u}$; project normally on that axis the centres of all the particles of the aggregate; determine the particle $A$ with the largest coordinate on the axis; place the centre of the new particle at unit distance from $A$ along the direction $\hat{u}$.
\end{itemize}
This growth process evolves stochastically because the orientation $\hat{u}$ changes at each newly added particles. A simulation result is shown in Fig. \ref{lambdainfty}c with a very large cluster of 500,000 particles in $2d$. The aggregate seems to tend toward a structure with 3 large branches making angles close to $2\pi/3$ with each other, but irregularities are still present in the form of small side-branches. This model was actually introduced in 1986 by R. Jullien for mimicking cluster growth \lq\lq by the tips" \cite{jullien1986new}. The author wrote at that time \lq\lq Although I have not yet found any experimental realisation, I consider [the model] instructive (...)". His numerical analysis suggested that the fractal dimension of the clusters was $1$, independently of the space dimension $d$ and possibly with logarithmic corrections in infinite $d$ \cite{jullien1986new}. The same rules were also re-obtained as a particular limit of an aggregation model where the diffusing particles interacted attractively within some range with the particles of the aggregate \cite{nicolas2016fractality,nicolas2017universal}. 

The above remarks allow us to conclude that the fractal dimension of the clusters formed by very short-lived particles (very high resetting rate)  is likely to be unity, {\it i.e.},
\begin{equation}\label{dflinfty}
d_f^{(\lambda=\infty)}=1,
\end{equation}
for Models A and B. 

\section{Numerical results}

We now test the above predictions with computer simulation experiments in $d=2$ dimensions by analysing the cases $\lambda<\infty$. Since long linear trajectories are extremely rare in Brownian motion and random walks, the limit $\lambda\rightarrow\infty$ is impossible to achieve in practice with standard particle dynamics. Nevertheless, this theoretical extrapolation is helpful to  gain insights on the outcomes at $\lambda\gg 1$ but finite, in the cases that are numerically tractable.

Numerical simulations are performed in discrete time and on a $L\times L$ square lattice with the seed at the centre. The launching circle has radius $R=L/3$. Most simulations are done with $R=160$ or $400$. We have used Eq. (\ref{pd}) with $\lambda\in[0,50]$, hence the largest value of $\lambda$ is $\gg 1$ but still keeps the death probability $p_d\ll 1$. Cluster related quantities are averaged over $30$ or $60$ realisations and a typical cluster size is $N\sim3000$. Larger clusters require a larger $R$ and take too long to grow at large $\lambda$. Unless otherwise indicated, the particles follow the standard unbiased random walk algorithm in discrete time, where  at each time step the walker jumps to a nearest-neighbour site, chosen with equal probability among all its neighbours. The n.n. sites belong to the von Neumann neighbourhood (hence, the diffusion coefficient is $D=1/4$). At the beginning of each walk, a total number of steps $t_{\ell}$ is chosen, where $t_{\ell}$ is a random variable distributed exponentially with mean $1/p_d$. The walk terminates either when it has performed $t_{\ell}$ steps without touching the aggregate, or when it touches the aggregate for the first time at a time lower than $t_{\ell}$. In the latter case, as in ordinary DLA, the criterion for aggregation is that at least one of the n.n. site of the walker is an aggregate site.

Figures \ref{drift}a and \ref{drift}b display two clusters obtained with Model A and B at a finite but large $\lambda$, respectively. The trajectory of the last particle added is indicated in each case. Clearly, these short and rather linear trajectories are drastically different from the random walks of ordinary DLA (Fig. \ref{drift}d). The clusters are also significantly less isotropic than DLA, as expected from Fig. \ref{lambdainfty}.

\subsection{Rare events vs. directed aggregation}

\begin{figure}
\centering
\includegraphics[width=0.85\linewidth]{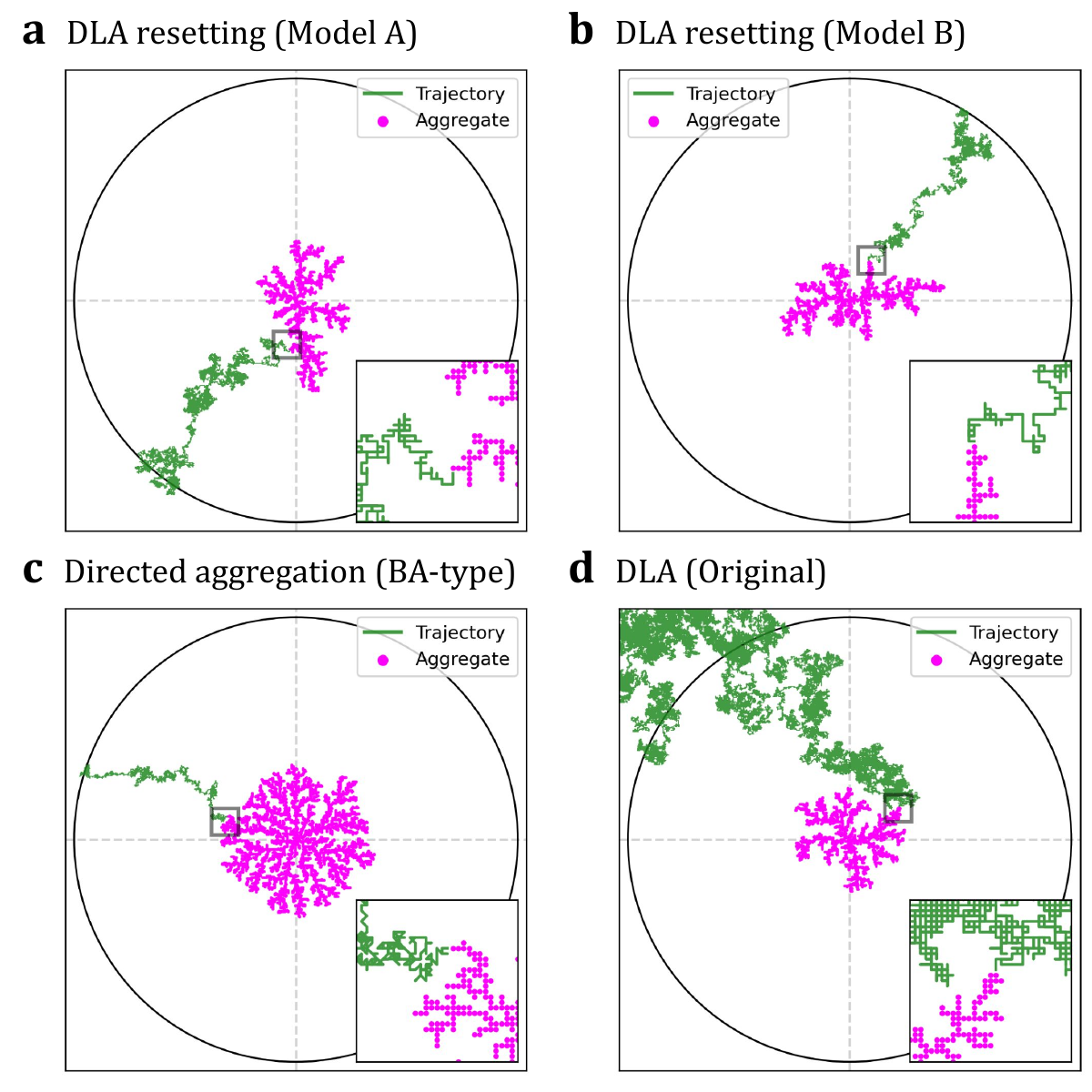}
\caption{(a) $2d$ aggregate (purple) formed in Model A with a re-scaled resetting rate $\lambda=40$. The green line is the trajectory of the last particle added. (b) Case of Model B, with $\lambda=50$ and the same colour code. (c)  Ballistic-like aggregation cluster produced via off-lattice diffusing particles with $4D=1$ and a drift velocity of magnitude $0.05$ always pointing toward the origin. 
The unbiased part of the random steps are displacement of unit length with uniform and random orientation in $[0,2\pi]$ (on the figure the position is rounded-off to the closest lattice site). (d) Ordinary DLA or $\lambda=0$. The insets are zooms near the aggregation zone.} \label{drift}
\end{figure}

Equation (\ref{dflinfty}) may seem surprising at first glance, as the clusters grown by the addition of particles that follow straight line trajectories are in general far from linear. A classical example of such processes is ballistic aggregation (BA), which involves particles that move in straight line in random directions and stick to the aggregate if it is found on the way \cite{vold1963computer}. Due to their linear nature, those trajectories are similar to the ones discussed in the previous section. However, the clusters in ballistic aggregation are known to be compact, {\it i.e.}, with fractal dimension $d_f=d$ \cite{meakin1983effects,ball1984causality,ramanlal1985theory}, in sharp contrast with the examples of Fig. \ref{lambdainfty}.

For comparing with our models at finite $\lambda$, we have simulated a ballistic-like aggregation process consisting of diffusing particles with a drift velocity directed toward the origin and of constant magnitude (see \cite{meakin1983effects,block1991aggregation} for related models). 
Figure \ref{drift}c confirms qualitatively the differences in cluster morphology obtained with the two types of drift: one which is explicitly imposed and the other one effective, resulting from a resetting constrain. Despite of the fact that the trajectories of the last particle added (green line) are similar in Figures \ref{drift}a, \ref{drift}b and  \ref{drift}c, the aggregate structures differ markedly: one is compact (Fig. \ref{drift}c) and the others more elongated and with a smaller number of large branches (Figs. \ref{drift}a and \ref{drift}b). This illustrates the specific impact that the large deviation regime has on the structure of the aggregate.

\subsection{Fractal dimension}

\begin{figure}
\centering
\includegraphics[width=0.68\linewidth]{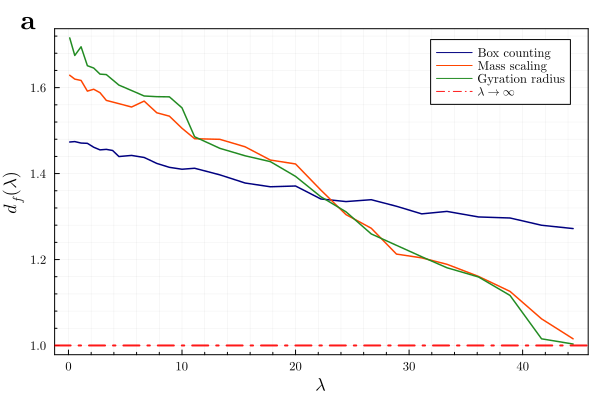}
\includegraphics[width=0.31\linewidth]{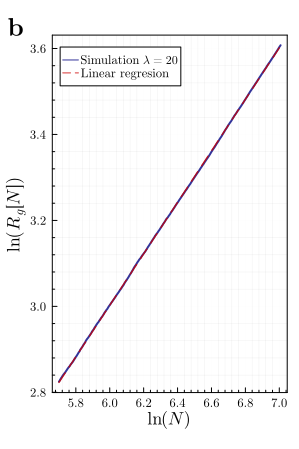}
\caption{(a) Fractal dimension of $2d$ clusters in Model A as a function of $\lambda$ by different measurement methods ($R=160$). (b) Scaling behaviour of the average gyration radius $R_g$ vs. the cluster size $N$ with $\lambda=20$.} \label{dimA}
\end{figure}

\begin{figure}
\centering
\includegraphics[width=0.68\linewidth]{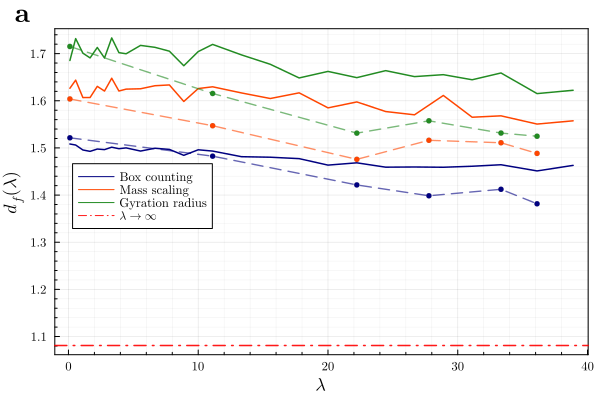}
\includegraphics[width=0.31\linewidth]{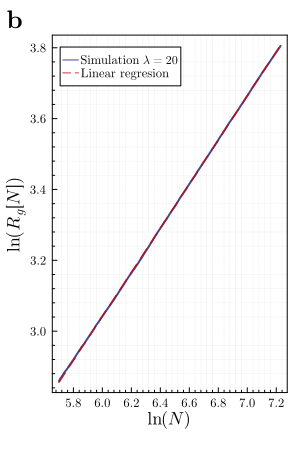}
\caption{(a) Fractal dimension of $2d$ clusters in Model B as a function of $\lambda$ by different measurement methods. The solid lines correspond to a launching circle with $R=160$ and the dashed ones to larger clusters, grown with $R=400$. The red horizontal line is the fractal dimension $1.08$ measured in aggregates obeying the rules of Model B ($\lambda=\infty$) with a finite $R$ of 2000 and $N\sim 1.5\ 10^4$.
 (b) Average gyration radius $R_g$ vs. the cluster size $N$ with $\lambda=20$.} \label{dimB}
\end{figure}

To determine the fractal dimension $d_f$ of the cluster and study its dependence on $\lambda$ in Models A and B, we have employed three standard methods, namely, the box-counting, mass scaling and gyration radius methods. In the first one,
the average number of square boxes of length $\epsilon$ needed to cover the structure scales as ${\cal N}(\epsilon)\simeq a \epsilon^{-d_f}$, with $a$ a constant. In the mass method, one uses the scaling relation $N(R)\simeq c R^{d_f}$ for the average number of aggregate particles at a distance less than $R$ from the origin, with $c$ a constant. The gyration dimension is defined through the relation obeyed by the average gyration radius of a cluster of $N$ particles, $R_g(N)=kN^{1/d_f}$, where $k$ is a constant (see Figs. \ref{dimA}b and \ref{dimB}b). For each $\lambda$, these quantities are averaged over the total number of realisations in order to estimate the corresponding fractal dimension. 

Figure \ref{dimA}a displays the fractal dimensions obtained as a function of the adimensional resetting rate $\lambda$, in Model A. The mass and gyration methods seem more reliable and consistent, as they give $1.62<d_f<1.71$ for $\lambda=0$, in agreement with the values reported in the literature on DLA (see \cite{nicolas2019universal} for a review). As predicted, we find that the fractal dimension consistently decays towards unity at large $\lambda$. 

Figure \ref{dimB}a shows the results for Model B, where the different methods are slightly less consistent between each other. In all cases $d_f$ also decreases with $\lambda$ from the DLA value, but, rather surprisingly, in a much slower way than Model A. We assume that this behaviour is due to two factors: one is the difference in the resetting protocols between Models A and B, and the other is the fact that the clusters of Model B reach their asymptotic $d_f$ at much larger sizes. To validate the latter hypothesis, we changed $R$ from $160$ to $400$ to grow larger clusters, and the results are represented by the dashed lines in Fig. \ref{dimB}. For a same $\lambda$, the larger clusters actually have a smaller fractal dimension. In any case, $d_f$ never increased from the DLA value toward the BA dimension of 2. Instead, the clusters became more linear with larger values of $\lambda$. 

The crossing between the curves observed in Figure \ref{dimA} but not in Figure \ref{dimB} seems to be related to the sensitivity of the box-counting method to low dimensional structures. The three methods provide relatively consistent measurements if the dimension of the clusters is above $d_f=1.3$, as it can be observed in Fig. \ref{dimB}. However, as the structure of the clusters become more linear, the box-counting method over-estimates the dimension leading to the crossing in Figure \ref{dimA}. The tendency of the box-counting method to under estimate $d_f$ for more compact clusters (such as DLA or BA) has been documented in the literature in the case of small clusters, as the ones analysed here. However, to the best of our knowledge, there are no systematic studies that explain the differences between methods (see Appendix A of ref. \cite{nicolas_thesis_2017}). 

One of the best approximate analytical results that describes the fractality of many aggregation models in $d$ dimensions ($d>1$) is the generalized Honda-Toyoki-Matsushita mean-field equation \cite{honda1986theory,matsushita1986generalization,hayakawa1986monte, nicolas2017universal},
\begin{equation}\label{honda}
d_f=\frac{d^2+\delta}{d+\delta},
\end{equation}
with $\delta=\eta(d_w-1)$, where
$\eta$ is a positive number associated with effects such as long range attractive interactions, screening or anisotropy (as introduced in the DBM \cite{niemeyer1984fractal}), and $d_w$ the fractal dimension of the particles' trajectories. For $\eta=1$ a good description is obtained for the fractal dimensions of the DLA-BA morphological transition. As the trajectories continuously transit from Brownian ($d_w=2$) to ballistic ($d_w=1$), $\delta$ goes from 1 to 0 in this case, hence $d_f$ increases from 1.67 (quite close to the \lq\lq exact" numerical value 1.71) to 2 in $2d$.  \cite{meakin1984levy, alves2006aggregation}. Obviously, relation (\ref{honda}) with any $\eta>0$ fails to explain why $d_f$ {\it decreases} in Figs. \ref{dimA} and \ref{dimB}, as the trajectories of the particles become more linear due to the resetting constrain. A starting point for a phenomenological description of this type would be to consider a modified $\delta$, of the form $\delta(d_w,\lambda)$, and fix $d_w=2$. The unknown function should then fulfil the conditions $\delta(2,0)=1$ and $\lim_{\lambda\to\infty} \delta(2,\lambda)=\infty$.

\section{Conclusion}
In this work we have studied fractal aggregation problems with particles subject to resetting, a type of process that models the finite lifetime of the building blocks diffusing around the aggregate. We have shown that the resulting clusters have rather uncommon features. In the conventional scaling theory of DLA and related models, the morphology of the cluster depends crucially on the scaling exponent describing the motion of the particles before aggregation. Here, however, in the large resetting rate limit, fractal growth is not controlled in the usual way by the fractal dimension of the trajectories of the aggregating elements but by a large deviation principle instead. This principle is well captured by a geometrical optics approximation of constrained Brownian motion. To our knowledge, the present contribution is one of the first to study an aggregation process of this kind.

In the short lifetime limit, our Model B provides a physical justification of a growth process \lq\lq by the tips" introduced some time ago \cite{jullien1986new} and which generates linear structures with non-trivial properties (Fig. \ref{lambdainfty}c). Some questions remain open regarding this process, for instance, whether the number of main branches 
always tends to 3 at large $N$ in $2d$ or whether other numbers are possible. Other properties such as the length distribution of the side-branches or the distances between them also deserve further study.

A very helpful method in the study of large deviation statistics is the design of efficient algorithms that are able to sample numerically rare trajectories or trajectories satisfying certain constrains \cite{majumdar2015effective,chetrite2015nonequilibrium}.  A classical example is the Doob's transform of the Langevin equation in the case of free Brownian motion, which leads to an effective Langevin equation generating Brownian paths starting at $x_0$ and that necessarily end at a given position $x$ at time $t_f$, with the correct statistical weights \cite{doob1957conditional}. In the derivation of effective Langevin equations for constrained diffusion problems, the constrains generally appear through an effective drift force which is time and space dependent.
The implementation of such an algorithm for a constrained DLA problem remains an open challenge. If it existed, one would be able to generate quickly many trajectories that stick anywhere on the aggregate with the correct probability. The difficulty in this context is that, unlike in the Doob's problem, the final position $x$ is variable and the time $t_f$ itself conditioned to be less than the particle lifetime $\tau$.

At finite but large $\lambda$, we have seen that our simulations exhibit some features of the $\lambda=\infty$ limit. Although the purpose of this work was not to explore the scaling properties of large-scale clusters, one would expect that the finite radius $R$ affects the results: under a finite lifetime constraint, it is easier for a diffusing particle to find the aggregate if it is launched after many previous particles (when the aggregate radius might be not so small compared to $R$) than at the beginning of the growth. This limitation could be the improved in future work, where the parameter $\lambda$ or $R$ would change over time as a function of the cluster characteristic length.

Finally, it would be interesting to explore the effects produced by resetting on other aggregation models. An quite natural candidate would be ballistic aggregation or related models that produce asymptotically compact clusters. Transitions to less compact fractals as observed here for DLA are expected, but whether the properties of such transition are generic or not, and whether they can be described within a unified framework are open questions.

\section*{Acknowledgements}
We thank Carlos E. L\'opez Natar\'en for technical computer support.
We acknowledge support from Ciencia de Frontera 2019 (CONACYT) Grant 263958.


\begin{thebibliography}{61}
\providecommand{\url}[1]{\texttt{#1}}
\providecommand{\urlprefix}{URL }

\bibitem{evans2020stochastic}
M.R. Evans, S.N. Majumdar and G. Schehr,  Journal of Physics A: Mathematical
  and Theoretical  \textbf{53} (19), 193001 (2020).

\bibitem{evans2011diffusion}
M.R. Evans and S.N. Majumdar,  Physical Review Letters  \textbf{106} (16),
  160601 (2011).

\bibitem{evans2011bdiffusion}
M.R. Evans and S.N. Majumdar,  Journal of Physics A: Mathematical and
  Theoretical  \textbf{44} (43), 435001 (2011).

\bibitem{majumdar2015dynamical}
S.N. Majumdar, S. Sabhapandit and G. Schehr,  Physical Review E  \textbf{91}
  (5), 052131 (2015).

\bibitem{reuveni2016optimal}
S. Reuveni,  Physical Review Letters  \textbf{116} (17), 170601 (2016).

\bibitem{chechkin2018random}
A. Chechkin and I. Sokolov,  Physical Review Letters  \textbf{121} (5), 050601
  (2018).

\bibitem{pal2017first}
A. Pal and S. Reuveni,  Physical Review Letters  \textbf{118} (3), 030603
  (2017).

\bibitem{eliazar2020mean}
I. Eliazar and S. Reuveni,  Journal of Physics A: Mathematical and Theoretical
  \textbf{53} (40), 405004 (2020).

\bibitem{maso2019transport}
A. Mas{\'o}-Puigdellosas, D. Campos and V. M{\'e}ndez,  Physical Review E
  \textbf{99} (1), 012141 (2019).

\bibitem{reuveni2014role}
S. Reuveni, M. Urbakh and J. Klafter,  Proceedings of the National Academy of
  Sciences  \textbf{111} (12), 4391--4396 (2014).

\bibitem{rotbart2015michaelis}
T. Rotbart, S. Reuveni and M. Urbakh,  Physical Review E  \textbf{92} (6),
  060101 (2015).

\bibitem{pal2019landau}
A. Pal and V. Prasad,  Physical Review Research  \textbf{1} (3), 032001 (2019).

\bibitem{kusmierz2014first}
L. Kusmierz, S.N. Majumdar, S. Sabhapandit and G. Schehr,  Physical Review
  Letters  \textbf{113} (22), 220602 (2014).

\bibitem{campos2015phase}
D. Campos and V. M{\'e}ndez,  Physical Review E  \textbf{92} (6), 062115
  (2015).

\bibitem{riascos2020random}
A.P. Riascos, D. Boyer, P. Herringer and J.L. Mateos,  Physical Review E
  \textbf{101} (6), 062147 (2020).

\bibitem{nagar2023stochastic}
A. Nagar and S. Gupta,  Journal of Physics A: Mathematical and Theoretical
  (2023).

\bibitem{gupta2014fluctuating}
S. Gupta, S.N. Majumdar and G. Schehr,  Physical Review Letters  \textbf{112}
  (22), 220601 (2014).

\bibitem{kang2022evolutionary}
Y.G. Kang and J.M. Park,  Journal of the Korean Physical Society  \textbf{81}
  (12), 1274--1279 (2022).

\bibitem{mercado2018lotka}
G. Mercado-V{\'a}squez and D. Boyer,  Journal of Physics A: Mathematical and
  Theoretical  \textbf{51} (40), 405601 (2018).

\bibitem{da2021diffusion}
T.T. da~Silva and M.D. Fragoso,  Journal of Physics A: Mathematical and
  Theoretical  \textbf{55} (1), 014003 (2021).

\bibitem{evans2022exactly}
M.R. Evans, S.N. Majumdar and G. Schehr,  Journal of Physics A: Mathematical
  and Theoretical  \textbf{55} (27), 274005 (2022).

\bibitem{grange2020non}
P. Grange,  Journal of Physics Communications  \textbf{4} (4), 045006 (2020).

\bibitem{magoni2020ising}
M. Magoni, S.N. Majumdar and G. Schehr,  Physical Review Research  \textbf{2}
  (3), 033182 (2020).

\bibitem{grange2020susceptibility}
P. Grange,  Journal of Physics Communications  \textbf{4} (9), 095018 (2020).

\bibitem{miron2021diffusion}
A. Miron and S. Reuveni,  Physical Review Research  \textbf{3} (1), L012023
  (2021).

\bibitem{grange2021aggregation}
P. Grange,  Journal of Physics A: Mathematical and Theoretical  \textbf{54}
  (29), 294001 (2021).

\bibitem{biroli2023extreme}
M. Biroli, H. Larralde, S.N. Majumdar and G. Schehr,  Physical Review Letters
  \textbf{130} (20), 207101 (2023).

\bibitem{meakin1998fractals}
P. Meakin, \emph{Fractals, scaling and growth far from equilibrium}, Vol.~5
  (Cambridge University Press, Cambridge, 1998).

\bibitem{sander2000diffusion}
L.M. Sander,  Contemporary Physics  \textbf{41} (4), 203--218 (2000).

\bibitem{Sander2011fractal}
L.M. Sander, in \emph{Mathematics of Complexity and Dynamical Systems}, edited
  by Robert~A. Meyers  (Springer New York, New York, NY, 2011), pp. 429--445.

\bibitem{niemeyer1984fractal}
L. Niemeyer, L. Pietronero and H.J. Wiesmann,  Physical Review Letters
  \textbf{52} (12), 1033 (1984).

\bibitem{witten1981diffusion}
T.A. Witten~Jr and L.M. Sander,  Physical Review Letters  \textbf{47} (19),
  1400 (1981).

\bibitem{witten1983diffusion}
T.A. Witten and L.M. Sander,  Physical Review B  \textbf{27} (9), 5686 (1983).

\bibitem{nicolas2019universal}
J. Nicol{\'a}s-Carlock and J. Carrillo-Estrada,  Scientific reports  \textbf{9}
  (1), 1120 (2019).

\bibitem{huang1987effects}
Y.B. Huang and P. Somasundaran,  Physical Review A  \textbf{36} (9), 4518
  (1987).

\bibitem{meakin1984levy}
P. Meakin,  Physical Review B  \textbf{29} (6), 3722 (1984).

\bibitem{meakin1983effects}
P. Meakin,  Physical Review B  \textbf{28} (9), 5221 (1983).

\bibitem{huang2001particle}
S.Y. Huang, X.W. Zou, Z.J. Tan and Z.Z. Jin,  Physics Letters A  \textbf{292}
  (1-2), 141--145 (2001).

\bibitem{block1991aggregation}
A. Block, W. Von~Bloh and H. Schellnhuber,  Journal of Physics A: Mathematical
  and General  \textbf{24} (17), L1037 (1991).

\bibitem{nakagawa1992extended}
M. Nakagawa, K. Kobayashi and H. Namikata,  Chaos, Solitons \& Fractals
  \textbf{2} (1), 1--10 (1992).

\bibitem{nicolas2016fractality}
J. Nicol{\'a}s-Carlock, J. Carrillo-Estrada and V. Dossetti,  Scientific
  reports  \textbf{6} (1), 19505 (2016).

\bibitem{turkevich1985occupancy}
L.A. Turkevich and H. Scher,  Physical Review Letters  \textbf{55} (9), 1026
  (1985).

\bibitem{basnayake2018extreme}
K. Basnayake, A. Hubl, Z. Schuss and D. Holcman,  Physics Letters A
  \textbf{382} (48), 3449--3454 (2018).

\bibitem{meerson2019large}
B. Meerson,  Journal of Statistical Mechanics: Theory and Experiment
  \textbf{2019} (1), 013210 (2019).

\bibitem{meerson2019geometrical}
B. Meerson and N.R. Smith,  Journal of Physics A: Mathematical and Theoretical
  \textbf{52} (41), 415001 (2019).

\bibitem{risken1989fokker}
H. Risken, \emph{The Fokker-Planck equation. Methods of solution and
  applications}   (Springer Science \& Business Media, Berlin, 1989).

\bibitem{zinn2004path}
J. Zinn-Justin, \emph{Path integrals in quantum mechanics}   (Oxford University
  Press, Oxford, 2004).

\bibitem{majumdar2005brownian}
S.N. Majumdar,  Current Science  \textbf{89}, 2076 (2005).

\bibitem{jullien1986new}
R. Jullien,  Journal of Physics A: Mathematical and General  \textbf{19} (11),
  2129 (1986).

\bibitem{nicolas2017universal}
J. Nicol{\'a}s-Carlock, J. Carrillo-Estrada and V. Dossetti,  Scientific
  reports  \textbf{7} (1), 3523 (2017).

\bibitem{vold1963computer}
M.J. Vold,  Journal of Colloid Science  \textbf{18} (7), 684--695 (1963).

\bibitem{ball1984causality}
R. Ball and T. Witten,  Physical Review A  \textbf{29} (5), 2966 (1984).

\bibitem{ramanlal1985theory}
P. Ramanlal and L. Sander,  Physical Review Letters  \textbf{54} (16), 1828
  (1985).

\bibitem{nicolas_thesis_2017}
J.R. Nicol\'as-Carlock, PhD thesis,  Benemérita Universidad Autónoma de
  Puebla, 2017, Available at
  \url{https://doi.org/10.6084/m9.figshare.24328759}.

\bibitem{honda1986theory}
K. Honda, H. Toyoki and M. Matsushita,  Journal of the Physical Society of
  Japan  \textbf{55} (3), 707--710 (1986).

\bibitem{matsushita1986generalization}
M. Matsushita, K. Honda, H. Toyoki, Y. Hayakawa and H. Kondo,  Journal of the
  Physical Society of Japan  \textbf{55} (8), 2618--2626 (1986).

\bibitem{hayakawa1986monte}
Y. Hayakawa, H. Kondo and M. Matsushita,  Journal of the Physical Society of
  Japan  \textbf{55} (8), 2479--2482 (1986).

\bibitem{alves2006aggregation}
S. Alves and S. Ferreira~Jr,  Physical Review E  \textbf{73} (5), 051401
  (2006).

\bibitem{majumdar2015effective}
S.N. Majumdar and H. Orland,  Journal of Statistical Mechanics: Theory and
  Experiment  \textbf{2015} (6), P06039 (2015).

\bibitem{chetrite2015nonequilibrium}
R. Chetrite and H. Touchette, in \emph{Annales Henri Poincar{\'e}}, Vol.~16
  (Springer, Berlin, 2015), pp. 2005--2057.

\bibitem{doob1957conditional}
J.L. Doob,  Bulletin de la Soci{\'e}t{\'e} Math{\'e}matique de France
  \textbf{85}, 431--458 (1957).

\end{thebibliography}
\end{document}